\documentclass{jps-cp}
\usepackage{txfonts} 

\usepackage{cite}
\usepackage{color} 
\usepackage{comment}

\title{Helical magnetism from sliding Luttinger liquids in hole-doped twisted bilayer WTe$_2$}

\author{Yung-Yeh \textsc{Chang}$^{1}$ and Chen-Hsuan \textsc{Hsu}$^{1,2}$}

\inst{$^{1}$Institute of Physics, Academia Sinica, Taipei 115201, Taiwan \\
$^{2}$Physics Division, National Center for Theoretical Sciences, Taipei 106319, Taiwan}

\recdate{\today}

\abst{We investigate carrier-mediated helical magnetism in hole-doped twisted bilayer WTe$_2$ coupled to localized moments. The itinerant carriers are described by a valleyless two-flavor sliding Luttinger liquid. Via the feedback mechanism of the helical order formed within individual channels, we show that the interchannel spin susceptibility mediates a Ruderman-Kittel-Kasuya-Yosida interaction between localized moments that can stabilize a coherent helical magnetic texture across the coupled-wire array. A spin-wave analysis of the ordered state allows us to estimate the magnon energies and helix-ordering scale. The magnon energies depend on the carrier interactions with representative values falling in the microwave regime, suggesting possible detection by magnetic-resonance techniques. The helical order partially gaps the carrier spectrum and renormalizes the spin-correlation and relaxation exponents. These results identify twisted bilayer WTe$_2$ as a tunable low-dimensional platform for realizing and probing carrier-mediated helical magnetism.
}

\kword{Hole-doped twisted bilayer WTe$_2$, sliding Luttinger liquid, Ruderman-Kittel-Kasuya-Yosida coupling, helical magnetism, spin-wave analysis}

\begin{document}

\maketitle

\section{Introduction}

Twisted layered nanostructures provide tunable platforms for strongly correlated fermions~\cite{Andrea-2020-NaPhys-moire, andrei-nat-rev-2021-moire,Mak:2022NatNanotech}. Their moiré potentials can organize carriers into one-dimensional channels, enhancing interaction effects and enabling coupled-wire physics~\cite{Hsu:2026}.
In particular, twisted bilayer WTe$_2$ exhibits experimental signatures consistent with coupled Luttinger-liquid behavior~\cite{2022-Nature-Pengjie-1DTLL-in-2DMoire,Sanfeng2023:tWTe2}. Theory predicts four-flavor channels in the electron-doped regime and valleyless two-flavor channels in the hole-doped regime~\cite{YiMing2024:tWTe2}. This carrier-dependent mode structure, together with twist-angle tunability~\cite{Lian:2024PRB_tWTe2}, provides a setting for exploring the interplay between one-dimensional correlations and magnetism. 

In this work, we focus on the hole-doped regime and investigate twisted bilayer WTe$_2$ coupled to localized moments. Unlike its valleyful counterpart, the hole-doped twisted bilayer WTe$_2$ system realizes a valleyless sliding Luttinger liquid, providing a comparatively direct setting. We derive the carrier-mediated Ruderman–Kittel–Kasuya–Yosida (RKKY) interaction from the exchange coupling between carriers and localized moments, following the coupled-wire calculations~\cite{YYC-2Dhelix-2025, YYC:2026arxiv} and similar analysis of isolated one-dimensional systems~\cite{Braunecker-PRL-2009-C13,Braunecker-2009,CHH-CNT-helix}. We show that the resulting carrier-mediated interaction can stabilize helical order of the localized moments at low temperatures. A spin-wave analysis provides estimates of the corresponding ordering scale and magnon energies, and shows that Coulomb interactions can further enhance the stability of the helical magnetism. 
We also determine the power-law exponents governing the local spin response above and well below the helix-ordering scale $T_{\rm hx}$.
The change of the exponent to the helix-renormalized one provides a signature of the reconstruction of the carrier sector by the ordered moments. These results establish twisted bilayer WTe$_2$ and related nanostructures as tunable low-dimensional settings for studying the interplay among strong correlations, carrier-mediated magnetism, and collective spin dynamics.

\section{Theoretical model and formulation }

We describe the parallel one-dimensional channels formed in twisted bilayer WTe$_2$ in the form of the sliding Luttinger liquid~\cite{Hsu:2023,YiMing2024:tWTe2}. The channels are labeled by an integer $m$, and the coordinate along each channel is denoted by $x$. 
For each channel, we introduce bosonic fields $\phi_{\nu,m}$ and $\theta_{\nu,m}$ for the charge and spin sectors, $\nu\in\{c,s\}$, respectively.
These fields satisfy
\begin{equation}
[\phi_{\nu,m}(x),\theta_{\nu',m'}(x')]
=
i\frac{\pi}{2}
\delta_{\nu,\nu'}\delta_{m,m'}
\mathrm{sgn}(x'-x),
\end{equation}
where $\phi_{\nu,m}$ describe density or spin-density fluctuations and $\theta_{\nu,m}$ are their conjugate fields.
For the carrier subsystem, the sliding Luttinger liquid describes a fixed point in which the channels are coupled through marginal forward-scattering interactions.

We impose periodic boundary conditions in the transverse direction and Fourier transform from $m$ to the transverse momentum $q_y$. The quadratic bosonic part of the Hamiltonian is then
\begin{equation}
H_{\mathrm{SLL}}
=\frac{1}{N_{\rm ch}}
\sum_{\nu\in\{c,s\}}\sum_{q_y}\int \frac{\hbar dx}{2\pi}\,
\left[
u_\nu (q_y) K_\nu(q_y)
|\partial_x\theta_\nu(q_y)|^2
+
\frac{u_\nu (q_y)}{K_\nu(q_y)}
|\partial_x\phi_\nu(q_y)|^2
\right] ,
\label{eq:H_SLL}
\end{equation}
where $N_{\rm ch}$ is the number of parallel channels, and $u_\nu(q_y)$ and $K_\nu(q_y)$ denote the $q_y$-dependent mode velocity and interaction parameter, respectively, in the $\nu$ sector. 
We use the reference carrier velocity $v_{\rm ch}$ to relate
the short-distance cutoff $a$ to the carrier bandwidth
$\Delta_a=\hbar v_{\rm ch}/a$. 
The charge-sector interaction parameter $K_c(q_y)$ is parametrized as
\begin{equation}
K_c(q_y)
=
K_c\big[1+\lambda_1\cos(q_y d)+\lambda_2\cos(2q_y d)\big],
\label{eq:Kc_qy}
\end{equation}
where $K_c<1$ describes repulsive intrachannel interactions~\cite{Giamarchi2003}, 
$\lambda_1$ and $\lambda_2$ specify the relative amplitudes of the first two transverse Fourier harmonics, and $d$ is the channel separation.
Our convention is reciprocal to that of
Refs.~\cite{Mukhopadhyay-sLL-PRB-2001,Carpentier-sLL-2001}; the Fourier coefficients here parametrize $K_c(q_y)$, not its inverse.
We retain general spin-sector parameters $K_s(q_y)=K_s$ and $u_s(q_y)=u_s$.

The coupling between the carrier spins and localized moments is described by
\begin{equation}
H_K =
J_K
\sum_{m } \int dx \; 
 {\bf s}_m (x) 
 \cdot 
{\bf S}_m (x) 
=
J_K
\sum_{m} \sum_{\mu, \sigma,\sigma'}\int dx \; 
\Bigg[ 
\frac{1}{2}
\psi_{m,\sigma}^\dagger(x)
\sigma^\mu_{\sigma\sigma'}
\psi_{m,\sigma'}(x) 
\Bigg]
S_m^\mu(x),
\end{equation}
where $J_K$ is the effective exchange coupling and $S_m^\mu$ is the $\mu$ component of the localized moment in channel $m$. The carrier  spin density, ${\bf s}_m$, of the $m$th channel is expressed in terms of the Pauli matrix $\sigma^\mu$ with $\mu \in \{x,y,z\}$ and fermion fields $\psi_{m,\sigma}$, which can be bosonized as in the literature~\cite{Hsu:2023,YiMing2024:tWTe2}.

Integrating out the carriers to order $J_K^2$ gives the RKKY interaction between localized moments,
\begin{equation}
H_{\mathrm{RKKY}}
=
\sum_{m,n}\sum_{\mu}
\int dx dx'\,
S_{m+n}^\mu(x)
J^{\mu}_{n}(x-x')
S_{m}^\mu(x') , 
\label{Eq:H_RKKY}
\end{equation} 
with the RKKY coupling scale proportional to the spin susceptibility of the carriers, $J^{\mu}_{n}(x-x')\propto  J_K^2 \chi^{\mu }_{n}(x-x')$. 
The susceptibility is defined through the connected carrier spin
correlation function,
\begin{equation}
\chi^{\mu}_{n}(x)
=
-\int_0^{\beta \hbar} \frac{d\tau} {4} \,
\sum_{\sigma_1 \sigma_2 \sigma_3 \sigma_4}
\left\langle
T_\tau 
\Big[ \psi_{m+n,\sigma_1}^\dagger(x,\tau)
\sigma^\mu_{\sigma_1\sigma_2}
\psi_{m+n,\sigma_2}(x,\tau) 
\Big]
\Big[ \psi_{m,\sigma_3}^\dagger(0,0)
\sigma^\mu_{\sigma_3\sigma_4}
\psi_{m,\sigma_4}(0,0) 
\Big]
\right\rangle_c,
\end{equation}
where $\beta=(k_BT)^{-1}$, $T_\tau$ denotes imaginary-time
ordering, and  $\langle\cdots\rangle_c$ denotes the expectation value associated with the carrier subsystem.
In the symmetry-preserving bare SLL state, the intrachannel ($n=0$) spin susceptibility is nonzero, whereas the interchannel ($n\neq 0$) susceptibility vanishes. 
In the spin-rotation-symmetric limit, this follows from the independent SU(2) symmetry of each channel in Eq.~\eqref{eq:H_SLL}. 

The nonzero intrachannel susceptibility mediates an RKKY interaction that favors a helical arrangement of the localized moments along each channel, with a wave vector selected by the dominant spin response near $q_x=\pm 2k_F$, where $k_F$ is the carrier Fermi wave vector. To analyze spin-wave excitations about this candidate texture, we locally rotate the spin coordinates so that the helical moments become uniformly aligned in the rotated frame. We then apply the Holstein-Primakoff transformation to express the spins in terms of magnon operators $a_m$ and $a_m^\dagger$. Retaining terms up to quadratic order yields the magnon Hamiltonian, which we diagonalize using a bosonic Bogoliubov transformation. At this stage, the RKKY interaction acts only within individual channels, leaving their relative helical phases undetermined.

However, this leading-order description, obtained by expanding to second order in $J_K$ about the bare SLL state, neglects the feedback of the magnetic texture on the carriers. 
To incorporate this feedback~\cite{Simon:2008},
we consider the helical exchange field generated by the ordered moments. This field introduces a sine-Gordon perturbation that, when relevant, partially gaps the carrier spectrum and mixes charge and spin degrees of freedom~\cite{Braunecker-PRL-2009-C13,Braunecker-2009,Hsu:2020}. 
We reevaluate the spin susceptibility in the reconstructed carrier state by transforming to combinations adapted to the helical perturbation and adopt a decoupled-sector approximation, neglecting the quadratic gradient terms that couple the gapped and gapless sectors. 
For a fixed helical texture, the carrier Hamiltonian including the exchange field no longer possesses the independent SU(2) symmetries of the bare SLL. The remaining gapless modes contain both charge and spin components, and their spin correlations must therefore be evaluated in the reconstructed state with interchannel Coulomb interactions included.

We denote the susceptibility in the presence of feedback by $\tilde{\chi}^{\mu}_n$.  In real space, the transverse components have the asymptotic form
$\tilde{\chi}^{x}_{n}(x) =
\tilde{\chi}^{y}_{n}(x)
\propto
-\cos(2k_F x)/{|x|^{2\overline K_n-1}},
$
corresponding to an enhanced response near longitudinal momenta $q_x=\pm 2k_F$. The power-law exponent is determined by the effective interaction parameter $\overline K_n$, which incorporates the modification of carrier correlations due to the helical feedback and partial gapping:
\begin{equation}
\overline{K}_n = \int_{-\pi/d}^{\pi/d}
\frac{dq_y}{(2\pi/d)}\,
\cos(nq_y d)
\left[
\frac{
u_c (q_y) K_c(q_y)+u_s/K_s
}{
u_s K_s+u_c  (q_y) /K_c(q_y)
}
\right]^{1/2}.
\label{eq:Kbar_n}
\end{equation}
For the subsequent evaluation of correlations and observables, we use representative parameters $K_s\approx 1$ and $u_s\simeq u_c(q_y)\simeq v_{\rm ch}$.

\section{Predicted observable features}

\subsection{Magnon energy scales }

The resulting interchannel RKKY coupling fixes the relative phases of the helices in different channels. For the phase-locked configuration with zero relative phase~\cite{YYC-2Dhelix-2025}, the classical spin texture takes the form
\begin{equation}
\mathbf{S}_m (x)
=
S
\left[
\cos(2k_Fx)\hat{\mathbf{x}}
+
\sin(2k_Fx)\hat{\mathbf{y}}
\right],
\label{Eq:xy-ansatz}
\end{equation}
where $S$ is the magnitude of the localized moment, and the helical wave vector $2k_F$ is selected by the extremum of the RKKY coupling. We then apply the spin-wave analysis to this phase-locked texture, using the RKKY couplings evaluated from $\tilde{\chi}^{\mu}_n$, to obtain the feedback-modified magnon spectrum.
The characteristic magnon energy extracted from the spectrum is thus set by the RKKY interaction, $\hbar\omega_0 \propto |\tilde{J}^x_{q_y=0}(2k_F)|$, where $\tilde{J}^\mu_{q_y}(q_x)$ denotes the Fourier component of $\tilde{J}^\mu_n(x) \propto \tilde{\chi}^{\mu}_n$. This energy scale provides an estimate of the stability of the helical magnetic state.

\begin{figure}[t]
    \centering
    \includegraphics[width=0.9\textwidth]{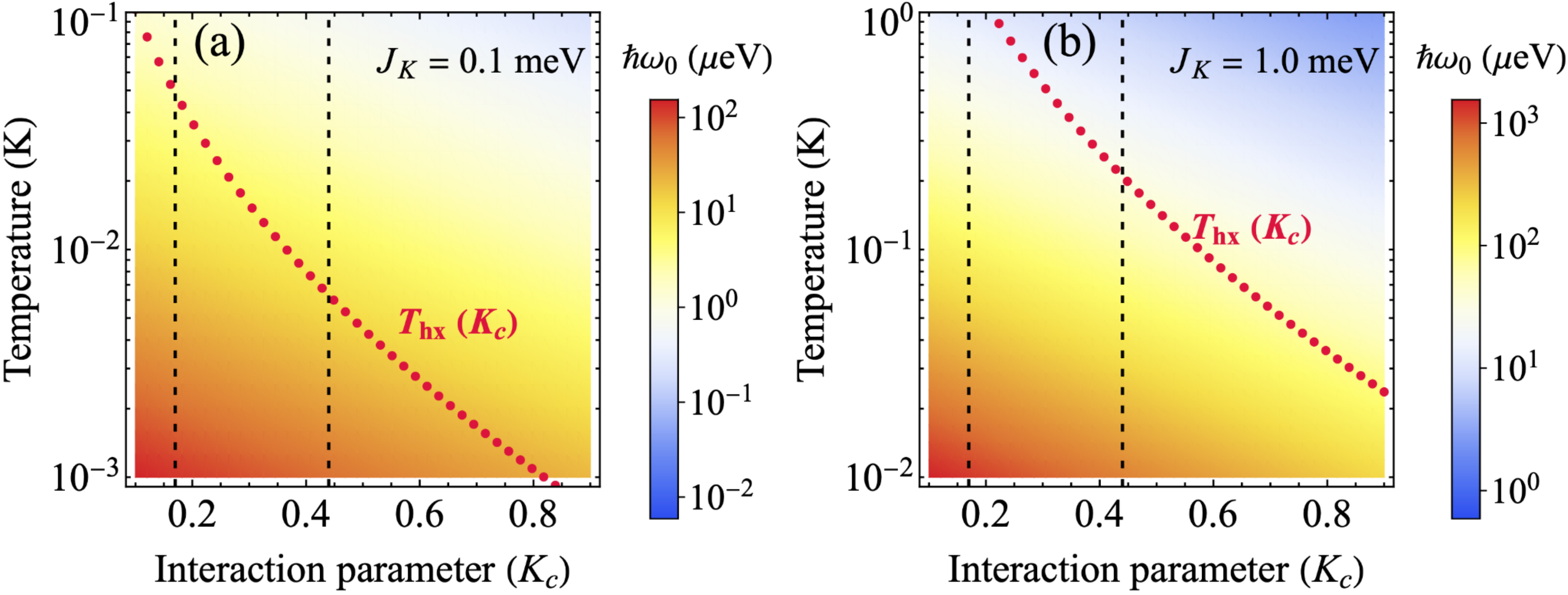}
    \caption{Magnon energy scale $\hbar\omega_0$ in the helically ordered coupled-wire system of hole-doped twisted bilayer WTe$_2$ coupled-wire system for (a) $J_K = 0.1$ meV and (b) $J_K = 1.0$ meV. The red dotted curves indicate the estimated helix-ordering scale $T_{\rm hx}$. The two black dashed lines mark the reported values of  $K_c = 0.17$ and  $K_c = 0.44$ in the  previous studies~\cite{Sanfeng2023:tWTe2,YiMing2024:tWTe2}. The parameters used here are  $\Delta_a = 5$~meV,  $v_{\rm ch} = 2 \times 10^4\, {\rm m/s}$, channel length $L = 0.5\, \mu{\rm m}$, number of channels   $N_{\rm ch} = 20$, and $\lambda_1 = \lambda_2 = 0.2$. 
     }
\label{fig:magnon}
\end{figure}

Figure~\ref{fig:magnon} shows representative estimates of magnon energies and ordering scale $T_{\rm hx}$. Repulsive carrier interactions enhance the RKKY coupling, raising magnon energies and making the helical magnetic state more robust. 
For the exchange couplings considered in Fig.~\ref{fig:magnon}, the estimated  $T_{\rm hx}$ values depend strongly on $K_c$, ranging from millikelvin to approximately kelvin scales.  
The colormap shows the magnon energies, with representative values in the microwave range motivating magnetic-resonance measurements as a probe of the helix.
An oscillating magnetic field could resonantly excite modes with nonzero magnetic-dipole matrix elements for the applied field. Electrical readout may be possible if these excitations produce a detectable change in carrier transport, by analogy with resistively detected magnetic resonance in semiconductors~\cite{Hirayama:2009SST}. Surface acoustic waves (SAW) offer another potential route~\cite{Jorge:2022APLReview,Jorge:2023PRL}, provided magnetoelastic coupling is appreciable and the acoustic and magnon modes satisfy the frequency, momentum, and symmetry requirements for coupling.
We next examine the consequences of the reconstructed carrier correlations for spin relaxation.

\subsection{Spin relaxation across the helical crossover }

The spin relaxation time $T_1$ provides a dynamical probe of the carrier-mediated helical state. The rate is determined by the itinerant carriers’ low-frequency transverse spin fluctuations:
\begin{equation}
\frac{1}{T_1T}
=
-\frac{J_K^2a^2k_B}{\hbar^3}
\lim_{\omega\rightarrow0}
\frac{{\rm Im} \, \chi_\perp^R(x=0,\omega)}{\omega},
\label{eq:T1_general}
\end{equation}
where $k_B$ is the Boltzmann constant, and
\begin{equation}
\chi_\perp^R(x,t)
=
-i\Theta(t)
\left\langle
\left[
s_m^+(x,t),s_m^-(0,0)
\right]
\right\rangle_c,
\qquad
s_m^\pm=s_m^x\pm i s_m^y .
\end{equation}
For repulsive carrier interactions, the dominant low-temperature contribution arises from spin-flip backscattering. For $T>T_{\rm hx}$, before helix formation, all gapless carriers described by Eq.~\eqref{eq:H_SLL} contribute. For $T\ll T_{\rm hx}$, the helical exchange field pins a linear combination of bosonic fields from the charge and spin sectors, whereas the orthogonal combination remains gapless. The contribution from the gapped sector is exponentially suppressed at low temperatures. The remaining gapless sector gives
$ {1}/({T_1T})
\propto
\left[1-m_{\rm hx}(T)\right]
T^{2\overline K-2}$, where $\overline K\equiv\overline K_{n=0}$ [see Eq.~\eqref{eq:Kbar_n}] is the renormalized interaction parameter  and $m_{\rm hx}(T)$ denotes the normalized helical order parameter, with $m_{\rm hx}(0)=1$. In our relaxation model, the reduction factor $1-m_{\rm hx}$, obtained from spin-wave analysis, weights the spin-flip contribution associated with thermally reduced helical order. Over the low-temperature fitting window, we parametrize the depletion as $1-m_{\rm hx}(T)\propto T^{2\alpha}$, with $\alpha$ determined by a numerical fit.

\begin{figure}[t]
    \centering
    \includegraphics[width=0.9\textwidth]{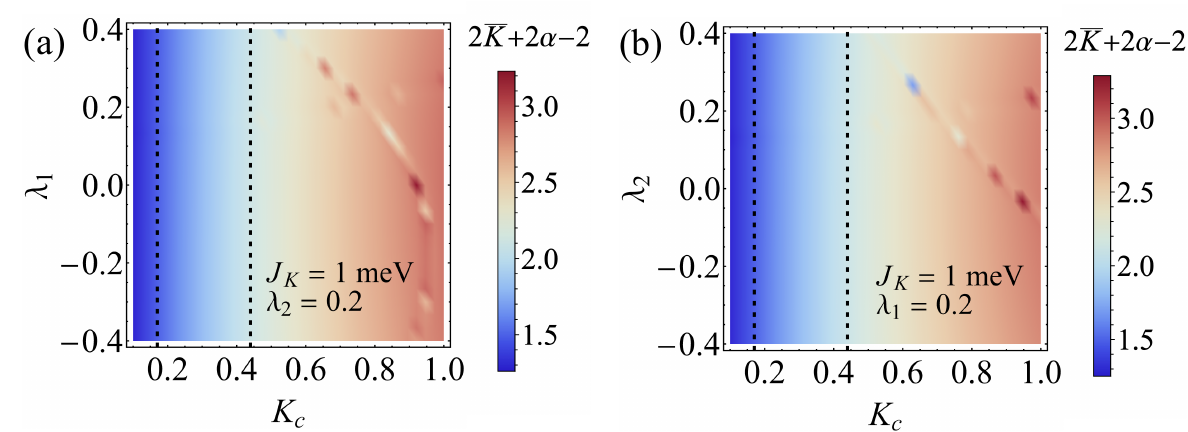}
    \caption{$1/(T_1T)$ exponent in the $T\ll T_{\rm hx}$ regime as a function of (a) $K_c$ and $\lambda_1$, and (b) $K_c$ and $\lambda_2$. For reference, the dashed lines mark $K_c=0.17$ and $0.44$. The remaining parameter values  are the same as in  Fig.~\ref{fig:magnon}.  
    } 
    \label{fig:gtilde_x}
\end{figure}

In summary,  the temperature dependence of $1/(T_1 T)$ in the regimes $T>T_{\rm hx}$ and $T\ll T_{\rm hx}$ reads
\begin{equation}
\frac{1}{T_1T}
\propto
\begin{cases}
T^{2g^x_0-2},
& T>T_{\rm hx},
\\[2pt]
T^{2 {\overline K}+ 2 \alpha  -2},
& T\ll T_{\rm hx},
\end{cases}
\qquad g^x_0 = \frac{1}{2}(K_c + K_s^{-1}). 
\label{eq:T1_summary}
\end{equation}
The absence of the additional valley sectors distinguishes Eq.~\eqref{eq:T1_summary} from the corresponding result for twisted bilayer graphene~\cite{YYC-2Dhelix-2025}. Thus, $1/(T_1T)$ exhibits two distinct asymptotic power laws above and well below $T_{\rm hx}$, providing a dynamical signature of spin-helix formation. In the immediate vicinity of $T_{\rm hx}$, the system is in a crossover regime that is not described by either asymptotic form. 

Above $T_{\rm hx}$, the scaling dimension governing the transverse $2k_F$ spin correlations is $g_0^x=(K_c+K_s^{-1})/2$. For $K_s=1$, the relaxation response scales as $1/(T_1T)\propto T^{K_c-1}$ and is therefore enhanced upon cooling for $K_c<1$ within the normal-state scaling regime. The independence of this exponent from $\lambda_1$ and $\lambda_2$ follows from the vanishing transverse averages of the cosine harmonics in Eq.~\eqref{eq:Kc_qy}. More general coupled-wire models can exhibit dependence on both interchannel interactions and the channel index~\cite{HCWang-TBG}.
In the fully noninteracting carrier limit, $K_c(q_y)=K_s=1$, the normal-state expression recovers the Korringa law, $1/(T_1T)=\mathrm{const}$.

Once the spin helix forms, the partial carrier gap modifies the contribution from the remaining gapless modes, while the reduction factor $1-m_{\rm hx}(T)$ introduces additional temperature dependence. Within the adopted relaxation model, the resulting low-temperature exponent is $2\overline K+2\alpha-2$, where $\bar K$ describes the carrier contribution and $\alpha$ characterizes the thermal depletion. As shown in Fig.~\ref{fig:gtilde_x}, this exponent depends
primarily on $K_c$, with weaker dependence on $\lambda_1$ and $\lambda_2$.
Taking $K_c\to1$ at fixed nonzero $\lambda_1$ or $\lambda_2$ does not eliminate charge interactions. Moreover, within the ordered-state model, a temperature-independent response  is not guaranteed by $K_c\to1$.

For a nanoscale device, resistively detected magnetic resonance may offer a practical probe of spin relaxation~\cite{Hirayama:2009SST}. Microwave excitation of an accessible magnetic mode at the energy scales shown in Fig.~\ref{fig:magnon} could perturb the spin system. The subsequent recovery of the electrical response may provide access to $T_1$, provided that the recovery is governed by spin relaxation.

\section{Discussion and conclusion}

We have investigated carrier-mediated helical magnetism in hole-doped twisted bilayer WTe$_2$ coupled to localized moments. Within our valleyless two-flavor sliding-Luttinger-liquid description, strong carrier correlations enhance the RKKY interaction, favoring helical order with a wave vector selected by the spin susceptibility. 
In a broader setting, interchannel hopping, which is absent from Eq.~\eqref{eq:H_SLL} but considered in coupled-wire descriptions of twisted bilayer WTe$_2$~\cite{YiMing2024:tWTe2}, can break the independent spin-rotation symmetries of individual channels. Spin-orbit coupling provides an additional source of spin anisotropy.
These effects are not included in the present analysis, but their influence on helical magnetism would be interesting to investigate in future work.

Spin-wave analysis provides estimates of the magnon energies and characteristic ordering scale, with representative energies in the microwave regime suggesting magnetic-resonance probes and, where appreciable magnetoelastic coupling is present, SAW-driven resonance.  Complementary to spectroscopic and magnetic-resonance measurements, Josephson interferometry in a superconductor--normal-metal--superconductor junction whose normal region hosts the helical magnetic state has been shown to provide an alternative transport-based probe of its spatially rotating magnetic texture~\cite{YYC:2026arxiv}. 

The helical order partially gaps the carrier spectrum and modifies the spin-correlation and relaxation exponents, yielding signatures distinct from those of valleyful coupled-wire systems~\cite{YYC-2Dhelix-2025}. Spin-relaxation measurements can therefore probe the feedback between correlated carriers and the magnetic order they mediate. Carrier density, twist angle, and screening can tune this interplay in a moiré material with experimentally established one-dimensional transport.

\section*{Acknowledgments} 

We acknowledge interesting discussions with  S.-F.~Lee and J.~Puebla. This work was financially supported by the National Science and Technology Council (NSTC), Taiwan (Grant No.~NSTC-114-2112-M-001-057 and Grant No.~115-2112-M-001-024), and Academia Sinica (AS), Taiwan  (AS-iMATE-114-12).

\bibliographystyle{jpsj}
\bibliography{LowDimHelix}

\end{document}